\documentclass{aa}
\usepackage{graphics}
\begin{document}

   \thesaurus{08     
              (08.03.2;  
               08.03.3;  
               08.03.4;  
               08.15.1)} 
   \title{The Core-Wing Anomaly of Cool Ap Stars
   \thanks{Based on observations obtained at the European
           Southern Observatory, Paranal, Chile (ESO programme
           No. 65.I-0644).}  }

   \subtitle{Abnormal Balmer Profiles}

   \author{Charles R. Cowley\inst{1} 
   \and S. Hubrig$^{2}$
   \and T.A. Ryabchikova$^{4,6}$
   \and G. Mathys$^3$
   \and N. Piskunov$^5$
   \and \newline P. Mittermayer$^6$}

\institute{Department of Astronomy, University of Michigan,
           David Dennison Building, Ann Arbor, Michigan 48109-1090, 
           USA \newline (cowley@umich.edu) 
           \and Astrophysikalisches Institut Potsdam, An der Sternwarte
            16, 14482 Potsdam, Germany (shubrig@aip.de) 
           \and  European Southern Observatory, Casilla 19001,
            Santiago 19, Chile (gmathys@eso.org)
           \and Institute of Astronomy, Russian Academy of Sciences,
            Pyatnitskaya 48, 109017 Moscow, Russia (ryabchik@inasan.rssi.ru)
           \and Uppsala Astronomical Observatory, Box 515, 751, 20
            Uppsala, Sweden (piskunov@astro.uu.se)
           \and Institute for Astronomy, University of Vienna,
            T\"{u}rkenschanzstrasse 17, 1180 Vienna Austria
            \newline (mittermayer@astro.univie.ac.at)}

   \offprints{Cowley}

   \date{Received 5 Dec. 2000; accepted }

   \maketitle

   \begin{abstract}

The profiles of H$\alpha$ in a number of cool Ap stars are
anomalous.  Broad wings, indicative of temperatures in the
range 7000-8000K end abruptly in narrow cores.  The widths
of these cores are compatible with those of dwarfs with
temperatures of 6000K or lower.  This profile has been known
for Przybylski's star, but it is seen in other cool Ap's.
The H$\beta$ profile in several of these stars shows a similar
core-wing anomaly (CWA).  In Przybylski's star, the CWA is probably
present at higher Balmer members. 
We are unable to account for these profiles within the context
of LTE and normal dwarf atmospheres.  We conclude that the
atmospheres of these stars are not ``normal.''  This is contrary
to a notion that has long been held.      
  
   \end{abstract}

%

\section{Introduction}

It has been widely assumed that the profiles of the Balmer lines in
chemically
peculiar (CP) stars are normal.  This does not mean that
small difficulties in getting
precise fits do not occur, but that generally speaking,
such problems are no more
serious for CP stars than those with normal compositions (cf. Barklem,
Piskunov,
\& O'Mara 2000a).
\begin{figure}
\resizebox{\hsize}{!}{\includegraphics{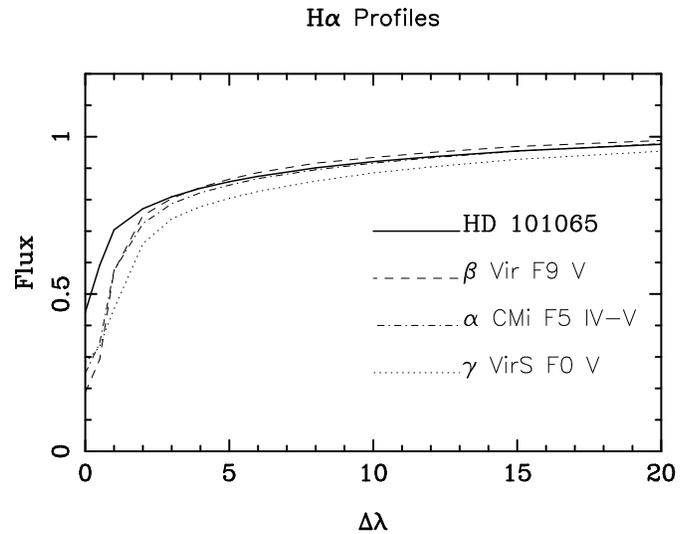}}
  \caption{H$\alpha$ profiles of HD 101065 and 3 F-dwarfs, plotted
   from data in Przybylski (1979).  Though Przybylski's resolution
   was much lower than that illustrated below, it was clear to him
   that the core-wing structure of HD 101065 was anomalous.}
\end{figure}
 
\begin{figure*}
\resizebox{12cm}{!}{\includegraphics{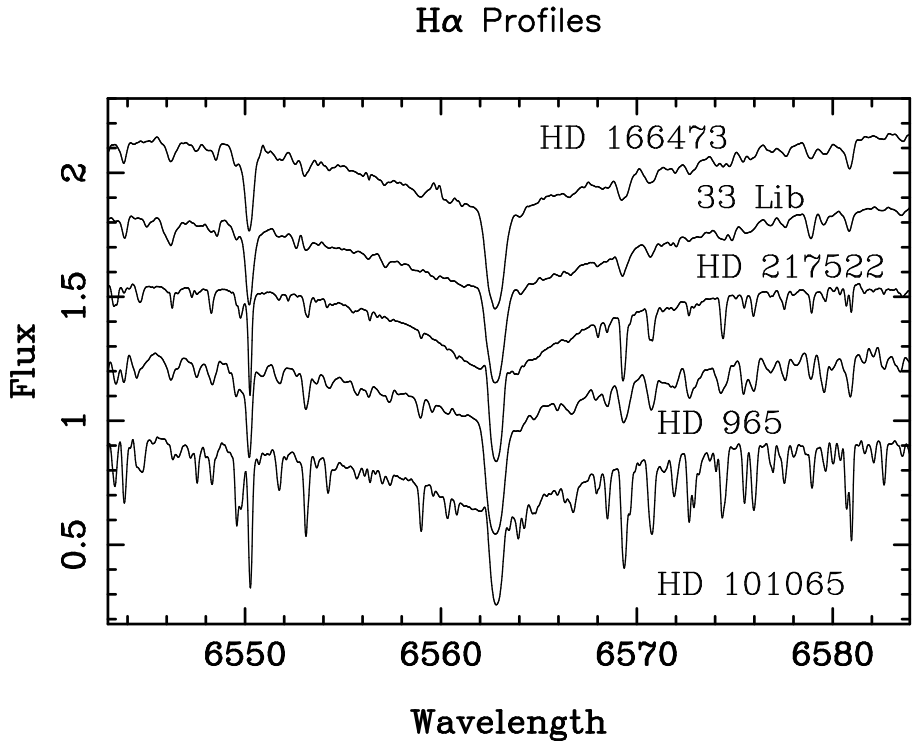}}
\hfill
\parbox[b]{55mm}{
\caption{H$\alpha$ profiles of 5 cool Ap stars showing the core-wing
anomaly (CWA).  
All spectra above that of HD 101065  have been displaced 
upward by 0.4, 0.6, 1.0, or 1.3 units.}
\label{Fig2}}
\end{figure*}

The strangest of all CP Stars, Przybylski's star
(HD 101065), was known to have
unusual Balmer profiles.  This was shown by Wegner (1976), and
Przybylski (1979) himself.  Figure 1 was constructed from a table in
Przybylski's paper, which appeared in one of the less accessible
journals.  It illustrates how the core of the H$\alpha$ profile in HD
101065 is markedly sharper than those of a sample of early to late
F-stars.  From the wing strengths alone, one might conclude that HD
101065 is a middle F-dwarf.  This, indeed, was the conclusion of Cowley
et al. (2000, henceforth {\bf  P1}) in a recent abundance analysis.
They adopted $T_{\rm e} = 6600$K, and $\log(g) = 4.2$, very close to
recent tabulations for mid-F dwarfs (cf. Drilling and Landolt 1999).

The authors of  {\bf P1} were aware of the unusual nature of the Balmer
lines in HD 101065, but had no solution to the riddle of the sharp
cores of H$\alpha$ and H$\beta$.  Both Przybylski and Wegner suggested
that these profiles might be understood in terms of an atmosphere that
was influenced by severe line blanketing.  Within the context of an LTE
calculation, we have been unable to reproduce both the deep, narrow
cores and the wings.  Several numerical experiments with blanketed
models were investigated in the preparation of {\bf P1}.  Even if the
boundary temperature were artificially lowered below the values in the
models discussed in that paper, it was not possible to reproduce the 
narrow, deep H$\alpha$ core.

\section{Balmer lines of other cool, magnetic CP stars}
 
The profiles of the Balmer lines of CP stars in the magnetic sequence
have not been extensively studied.   
It is very difficult to reduce high-resolution
echelle spectra in such a way that the profiles of the Balmer lines
are reliable.
These objects are generally more difficult
to study than their non-magnetic congeners, the Am and HgMn stars.
Their surfaces are known to show spatial inhomogeneities (abundance
patches), so the meaning of any traditional abundance study is
uncertain.  Much of the recent spectroscopic work on these stars has
been concerned with spatial mapping of the chemical inhomogeneities, or
line identifications.  
 
We have recently examined high resolution spectra of several cool CP
stars
in the magnetic sequence, and find some have deep cores at H$\alpha$
and H$\beta$ similar to those of HD 101065.  Figure 2 shows profiles 
(from top to bottom) of
HD 166473 (CoD -37 12303), HD 137949 (33 Lib), 
HD 217522 (CoD -45 14901), HD 965 (BD -0 21), and 
HD 101065 (CoD -46 5445). Spectra
of first two stars were obtained by PM at South Africa Astronomical
observatory (SAAO) using the fiber-fed  GIRAFFE echelle spectrograph
with a resolution of 35000. These spectra were reduced with standard
IRAF package,
but special efforts were taken to make a proper continuum level in
echelle orders with hydrogen lines.
For other three stars the spectra were obtained by
SH at ESO  using
the UV-Visual Echelle Spectrograph of the VLT.  The original
resolution is
estimated to be 110000.  The spectra were  mildly filtered, and
normalized by CRC using Michigan software, but the sharp cores are
plainly visible in the unnormalized spectra.

At the moment, it is not known whether
HD 965 is a rapidly oscillating Ap star (roAp), 
but other four stars illustrated are known
roAp's.

A CWA is present in H$\beta$ for some of the stars of Figure 2, as
shown in Figure 3.

\begin{figure}
\resizebox{\hsize}{!}{\includegraphics{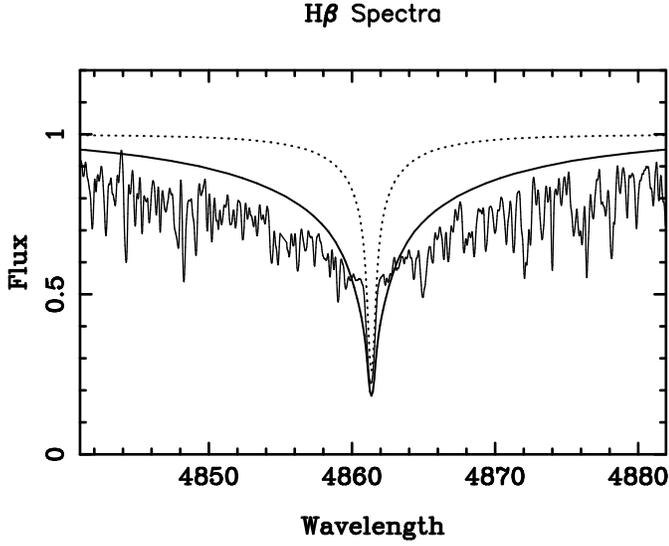}}
\caption{H$\beta$ profile in HD 965.  The smooth curves are from
the Kurucz (1994) library of profiles for normal abundances with
Vturb = 0.  The dashed curve is for $T_e = 5500$K, 
the solid curve for $T_e = 7000$K.  Both profiles are for
$\log(g) = 4.0$.  A high point in the stellar spectrum 
near $\lambda$4855 caused by hot pixels has
been artificially removed.}  
\label{Fig3} 
\end{figure}

The CWA is often less
pronounced at H$\beta$ than H$\alpha$.  In HD 101065, the anomaly
persists into the higher series members, as was noted by Wegner (1976).
In HD 217522, one can see already (Fig. 2) 
in H$\alpha$, a possible transition
case, where there appears a kind of intermediate-near wing outside the
core.  By H$\beta$, the CWA is hardly noticable in HD 217522.  
 
van't Veer and Megessier (1996) studied H$\alpha$ and H$\beta$ profiles
of the Sun, Procyon, and the cool Am stars 63 Tau and 
$\tau$ UMa (cf. their Fig. 5).  More
recently, Gardiner et al. (1999) discussed  H$\alpha$ profiles in a
dozen normal stars (including the sun) with effective temperatures
between 5777 K and 9940 K.   Their spectra 
do not have the sharp break in
slope that can be seen in our Fig.2. 
The recent study of the roAp star HR 3831 (HD 83332) by Baldry and 
Bedding (2000) may not have had sufficient resolution to reveal a CWA.
We tentatively conclude the
phenomena is limited to the magnetic sequence of CP stars.  It remains
to be seen if it is also limited to the roAp's.
 
The best studied non-roAp star is $\beta$ CrB. This star has similar
atmospheric parameters to the roAp star $\gamma$ Equ.
The spectrum of $\beta$ CrB was obtained at McDonald observatory with a
resolution of 56000. We display it below, along with a spectrum of 
$\gamma$ Equ from the same observational run at SAAO
in which the upper two spectra of Fig. 2 were obtained.

\begin{figure}
\resizebox{\hsize}{!}{\includegraphics{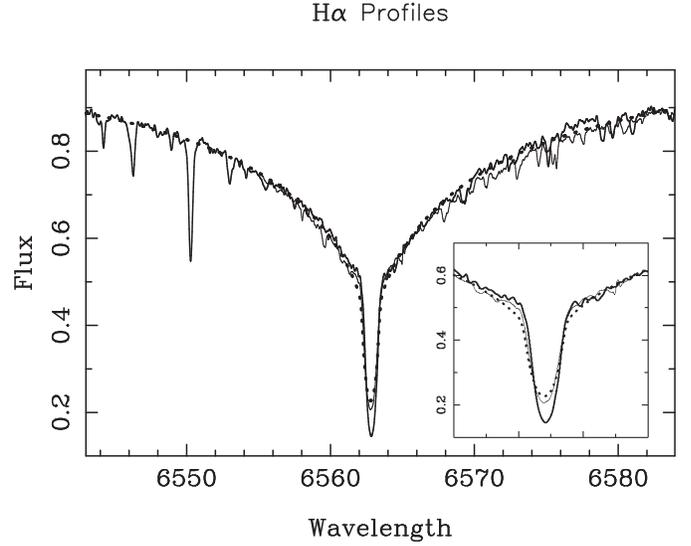}}
\caption{Observed H$\alpha$ profiles for $\gamma$ Equ (thick line), $\beta$ CrB,
(thin line) and a calculated profile (dotted) for $T_e = 7800$K, 
$\log(g) = 4.2$.   The inset shows the same data, which has been plotted
at a different scale to illustrate the near fit between the theoretical
calculations and the observations.  
}
\label{Fig4}
\end{figure}

We compare the H$\alpha$ lines for both stars with a theoretical line
profile calculated by spectrum synthesis code SYNTH, which incorporates
the latest version of hydrogen line broadening (Barklem, Piskunov
\& O'Mara 2001). The comparison is shown on Fig.4. Note the following: 
\begin{itemize}
\item[$\bullet$]{A break in the slope of H$\alpha$ line profile is similar for
      both roAp star $\gamma$ Equ and non-roAp star $\beta$ CrB, but
      it is less extreme than for the stars of Fig. 2.}
\item[$\bullet$]{This break is also present on theoretical line profile: the slow
      decaying wings of a hydrogen line are formed by energy level
      perturbations by electrons and protons. The quasi-static
      approximation used to compute the Stark profile predicts the
      very narrow core. The additional thermal Doppler broadening is
      virtually unnoticeable everywhere except in the line core. Thus,
      the core has a Doppler shape reflecting the temperature and
      velocity field of the top atmospheric layers.}
\item[$\bullet$]{It is prudent to say that the break in the slope is only,
      ``marginally'' different from theoretical predictions.}
\end{itemize}
 
The stars of Figure 2  are all cooler than
$\beta$ CrB and $\gamma$ Equ.  But at least one cool roAp, HD 122970
has an H$\alpha$ profile that is well fit by theory.  Possibly 
the roAp phenomenon may be a necessary condition for the CWA,
but not a sufficient one. 

\section{Interpretation}

We note that a core-wing structure develops in normal dwarfs 
with temperatures between 7000 and 8000K.  This is illustrated
in Figure 5, with the profiles taken from Kurucz (1994).
This structure may account for the reasonable fit shown in
Figure 4.  It is {\it not} what we are calling the CWA shown
in Figures 2 and 3.  Those stars are cooler than $\beta$ CrB
and $\gamma$ Equ, and have temperatures closer to
7000 than 8000K. Their H$\alpha$ profiles have a much 
more abrupt change in the slope between
the wing and the core than is seen in normal stars or the 
calculations of Figure 5.   

\begin{figure}
\resizebox{\hsize}{!}{\includegraphics{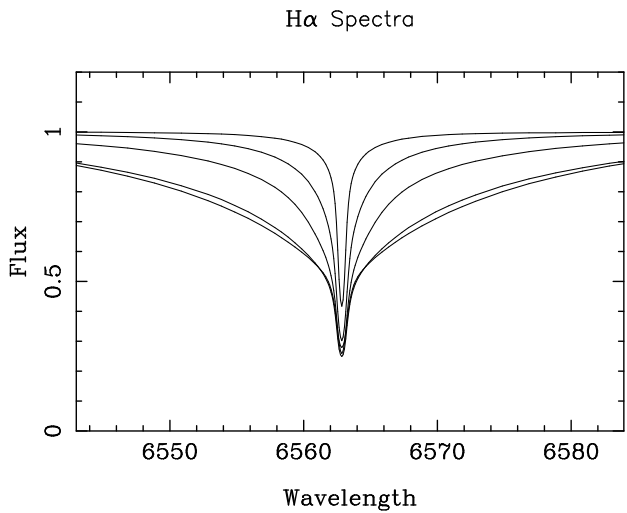}}
\caption{Calculated H$\alpha$ profiles for normal dwarfs
($\log(g) = 4.0$) with temperatures of 5000 (weakest line),
6000, 7000, 8000, and 9000K. 
}
\label{Fig5}
\end{figure}

Deep cores in an LTE calculation are  usually made by a low boundary
temperature.  In the case of hydrogen, the high-excitation $n = 2$
level becomes rapidly depopulated as the temperature drops, so that the
cores are relatively insensitive to a lower boundary temperature.   We
have attempted arbitrarily adding a cool, dense layer to the top of our
atmospheres.  In LTE, low-excitation metal lines become unrealistically
strong well before the H$\alpha$ core becomes as deep as in the
observations.
 
We have attempted to reproduce the CWA by arbitrary
modifications of the temperature distribution that might be attributed
to convection, or convection in a magnetic, pulsating atmosphere.
These attempts were completely unsuccessful.  For the
present, we assume that only a fully non-LTE calculation can
be expected to reproduce the observed profiles.

We have also performed non-LTE calculations using the 
MULTI code (Carlsson 1992)
for the first three Balmer lines for two model atmospheres with T$_{\rm
eff}$ = 6000 and 8000\,K. The resulting profiles have slightly (less
than
5\%) deeper cores with largest effect in H$\alpha$, but we find no
difference in the width of the core.  The potential effect of the
partial redistribution during scattering in the line cores 
remains to be investigated.

 
It now seems plain that the atmospheres of magnetic CP stars cannot be
considered normal, even setting aside the well known surficial chemical
inhomogeneities (patches).   Aside from the sharp cores of H$\alpha$
and H$\beta$, there is the ionization phenomena between the second and
third spectra of the lanthanide rare earths.   This problem was
discussed by Ryabchikova and her colleagues (cf.  Ryabchikova et
al. 2000, Gelbmann et al. 2000), and also in  {\bf P1}.  Essentially,
abundances from the third spectra, primarily of Pr III and Nd III are
an order of magnitude greater than those of the second spectra of these
elements.

It is unlikely the ionization anomaly  can be due to the neglect of
hyperfine structure
in Pr III or odd-N isotopes of Nd III, because the lines of these
spectra are relatively
weak in some of the roAp stars (cf. Ryabchikova et al. 2000, HD 122970
and 10 Aql).
Moreover, the same ions can be observed in non-roAp stars and do not
show the same
anomaly (Ryabchikova et al. 2001, Weiss et al. 2000).
 
Our partial analysis of HD 965 suggests that the same kind of 
ionization anomaly exists between Fe I and II.  The effective
temperature of a model which makes the abundances from Fe I and II
lines equal is 8000 -  9000K, while the color temperature of the 
star is in the range 7000 - 7500K.  A low surface gravity
palliates this somewhat, but no realistic gravity will reconcile
the ionization and color temperatures.

We are left with distinct
indications of two temperatures in models
whose structures are quite uncertain.  
The core depth of the early Balmer
lines argues for their formation in a region that covers most of the
star.  If, indeed, a high layer of gas overlies the photospheres of
these stars as in the model by Babel (1992), it could play an important
role in the origin of the chemical anomalies of these magnetic Ap
stars.
 
\begin{acknowledgements}

CRC thanks many colleagues for helpful conversations regarding
HD 965 and the CWA.  He thanks Dr. Saul J. Adelman for sending
Balmer profiles of several cool Ap stars, and C. van't Veer-Menneret
for commenting on profiles of Am-Fm stars.

\end{acknowledgements}

\end{document}